\documentclass[preprint,times,12pt]{elsarticle}

\usepackage{amssymb}

\newcommand{\etal}{\textit{et al.}}
\newcommand{\PRL}{Phys.~Rev.~Lett.~}
\newcommand{\APP}{Astropart.~Phys.~}
\newcommand{\ICRC}{\rm International~Cosmic~Ray~Conference}
\newcommand{\NIMPA}{Nucl.~Instrum.~Meth.~Phys.~Res.~A~}
\newcommand{\Cheren}{\rm Cherenkov}

\newcommand{\bvec}[1]{\mbox{\boldmath $#1$}}

\journal{Astroparticle Physics}

\begin{document}

\begin{frontmatter}



\title{Energy Spectrum of Ultra-High Energy Cosmic Rays\\ Observed with the Telescope Array\\ Using a Hybrid Technique}


\author[1]{T.~Abu-Zayyad}
\author[2]{R.~Aida}
\author[1]{M.~Allen}
\author[1]{R.~Anderson}
\author[3]{R.~Azuma}
\author[1]{E.~Barcikowski}
\author[1]{J.W.~Belz}
\author[1]{D.R.~Bergman}
\author[1]{S.A.~Blake}
\author[1]{R.~Cady}
% ...

\address[1]{High Energy Astrophysics Institute and Department of Physics and Astronomy, University of Utah, Salt Lake City, Utah, USA}
\address[2]{University of Yamanashi, Interdisciplinary Graduate School of Medicine and Engineering, Kofu, Yamanashi, Japan}
\address[3]{Graduate School of Science and Engineering, Tokyo Institute of Technology, Meguro, Tokyo, Japan}
% ...

\cortext[cor1]{Corresponding author. Tel./fax: +81-4-7136-5157.}

\begin{abstract}
　We measure the spectrum of cosmic rays with energies 
greater than $10^{18.2}$ eV with the Fluorescence Detectors
(FDs) and the Surface Detectors (SDs) of the Telescope 
Array Experiment using the data taken in our first 2.3-year 
observation from May 27 2008 to September 7 2010. A hybrid air shower 
reconstruction technique is employed to improve accuracies 
in determination of arrival directions and primary energies of 
cosmic rays using both FD and SD data.
The energy spectrum presented here is in agreement
with our previously published spectra and the HiRes results.
\end{abstract}

\begin{keyword}
    Ultra-high energy cosmic rays \sep Telescope Array \sep hybrid spectrum


    \end{keyword}

\end{frontmatter}


\section{Introduction}
\begin{figure}[!htbp]
    \begin{center}
        \includegraphics[width=\columnwidth]{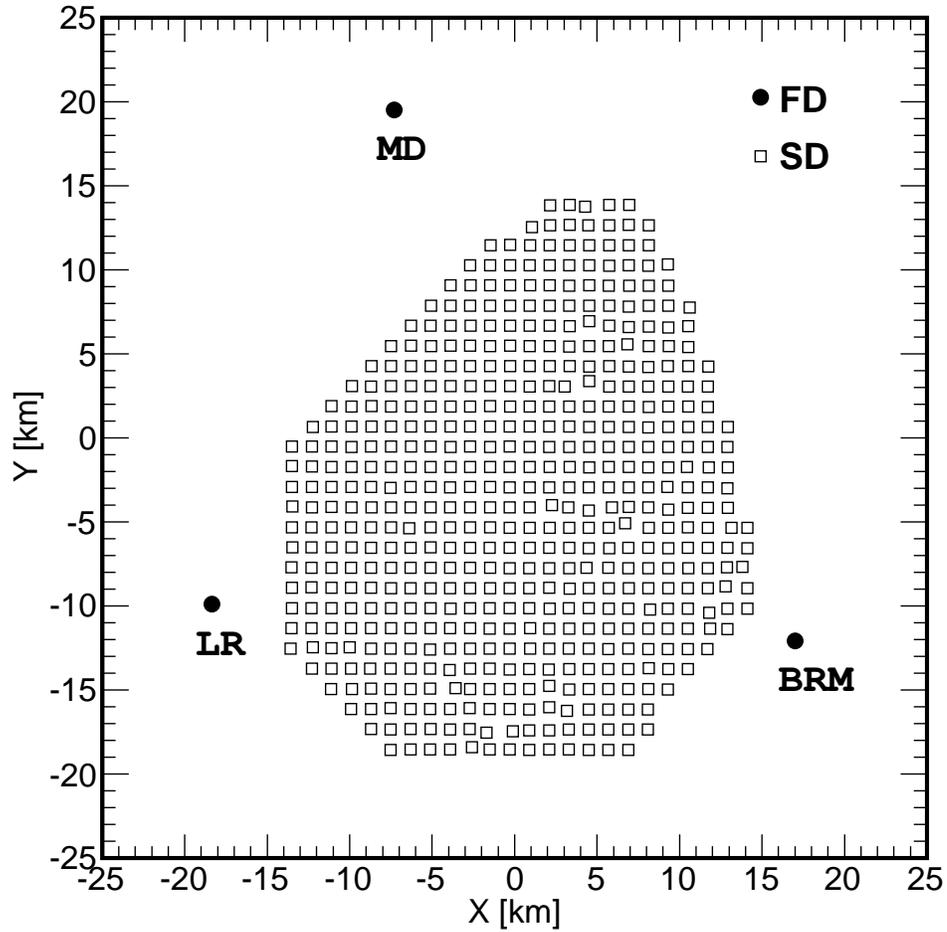}
    \end{center}
    \caption{The layout of the Telescope Array in Utah, USA. Open squares denote the 507 SDs. 
    The three filled circles denote the BRM, LR and MD FD telescope stations. 
    The horizontal (West-East) and vertical (South-North) axes indicate the locations of the TA detectors relative to the center of the site in km.
    }
    \label{fig:TA}
\end{figure}

The Telescope Array (TA) is the largest detector of ultra-high energy cosmic rays (UHECRs) in the northern hemisphere (see Figure~\ref{fig:TA}). It is designed to explore the origin of UHECRs and the mechanisms of production, acceleration at the sources, and propagation in the inter-galactic space.

The TA ~\cite{bib:TA1,bib:TA2} consists of 38 fluorescence detectors (FDs) and an array of 507 surface detectors (SDs). 
The FDs measure longitudinal development and primary energies of air showers in the atmosphere from the amounts of light emitted by atmospheric molecules excited by charged particles in the showers \cite{bib:TAFD}. 
The SDs measure arrival timings and local densities of the shower particles at the ground. The arrival direction and primary energy of an air shower in SD is determined from the relative timing differences of particle arrivals between SDs, and from the lateral distribution of local particle densities around the shower core, respectively \cite{bib:SDSpec}.
The advantage of FD is that air shower energies can be determined calorimetrically knowing the fluorescence yield, which is the amount of lights emitted by air molecules per total energy losses of charged particles in the showers. However there is a rather large uncertainty in arrival directions of cosmic rays determined with FD in monocular mode, in which time differences between signals of the photo-tube pixels with small angular separations are used.

A {\it hybrid} reconstruction technique, using the timing information of an SD at which air shower particles hit the ground, solves the problem. Our Monte-Carlo study shows that the inclusion of SD timing in FD monocular reconstruction significantly improves the accuracy in the determination of shower geometry (a similar method has been used in The HiRes-MIA~\cite{bib:HiResMIA} and the Pierre Auger Observatory ~\cite{bib:PAO-E}. 
The aim of this paper is to describe in full detail of our hybrid reconstruction method, and discuss the energy spectrum of ultra-high energy cosmic rays derived from this with improved accuracies in arrival directions and primary energies. Another advantage of our strategy is that the aperture of the detector can be simply calculated from that of the SD, which is almost determined geometrically.

This technique is also important to determine the composition of primary cosmic rays. Here, the FD's measure the shower development maximum in the atmosphere, X$_{\rm max}$, which is a parameter sensitive to the mass composition. Since this measurement is very sensitive to the shower geometry reconstruction, the hybrid technique's improved geometrical accuracy is important. The present work on the spectrum sets the stage for subsequent publications on primary composition using the same technique.

This paper is organized as follows.  We describe the TA detector in Section~\ref{sec:TA}.  The hybrid reconstruction method is given in Section~\ref{sec:analysis}. Section~\ref{sec:MC} explains air shower MC simulation and detector MC simulation.  We compare the distributions of data and MC in Section~\ref{sec:comparison}, and present the energy spectrum in Section~\ref{sec:result}.  The conclusion is described in Section~\ref{sec:summary}.

\section{The TA detectors} 
\label{sec:TA}
The TA site is located in Millard County, Utah, USA. The SD array covers an area of about 700~km$^2$. 
Each of the 3-m$^2$ SDs includes two layers of plastic scintillators wrapped with Tyvek reflective sheets in a stainless steel box. 
Scintillation photons produced by the passage of charged particles in air showers through scintillators are collected by a one-inch-diameter PhotoMultiplier Tube (PMT) for each layer.
The duty cycle of the SD is nearly 100\%.
Full details on the SDs can be found in ~\cite{bib:TASD}.

The TA FDs are installed in three stations (Black Rock Mesa [BR], Long Ridge [LR], and Middle Drum [MD]), which overlook the surface array. 
Each station contains 12 or 14 telescopes (12 at BR, 12 at LR and 14 at MD), observing 3$^\circ$ to 31$^\circ$ in elevation, and 108$^\circ$  for BR and LR and 120$^\circ$ for MD in azimuth.
The 14 MD telescopes are refurbished HiRes-1 detectors \cite{bib:TAMDSpec}.
The telescopes are operated on clear, moonless nights.
Each telescope collects and focuses ultraviolet fluorescence light emitted by nitrogen molecules in the wake of the extensive air showers using a spherical mirror of 6.8~m$^2$ effective area. 
This light is detected by cameras which consist of 256 PMTs (HAMAMATSU; R9508).
The PMT signals are sampled by FADC-based electronics with an effective rate of 10~MHz and a 14-bits dynamic range.
Detailed description of DAQ system are presented elsewhere ~\cite{bib:TAFD,bib:SDF,bib:TFCTD}.


We have a steerable mono-static LIDAR system \cite{bib:atm-atten} at the BR site to monitor atmospheric transparency by measuring backscattered light from a dedicated 355-nm Nd:YAG laser.

\section{Hybrid Reconstruction and Event Selection}
\label{sec:analysis}

The process of analysis consists of four steps: PMT selection, shower geometry reconstruction, reconstruction of longitudinal shower profile and quality cuts.

\begin{figure}[!htbp]
    \begin{center}
        \includegraphics[width=0.47\columnwidth]{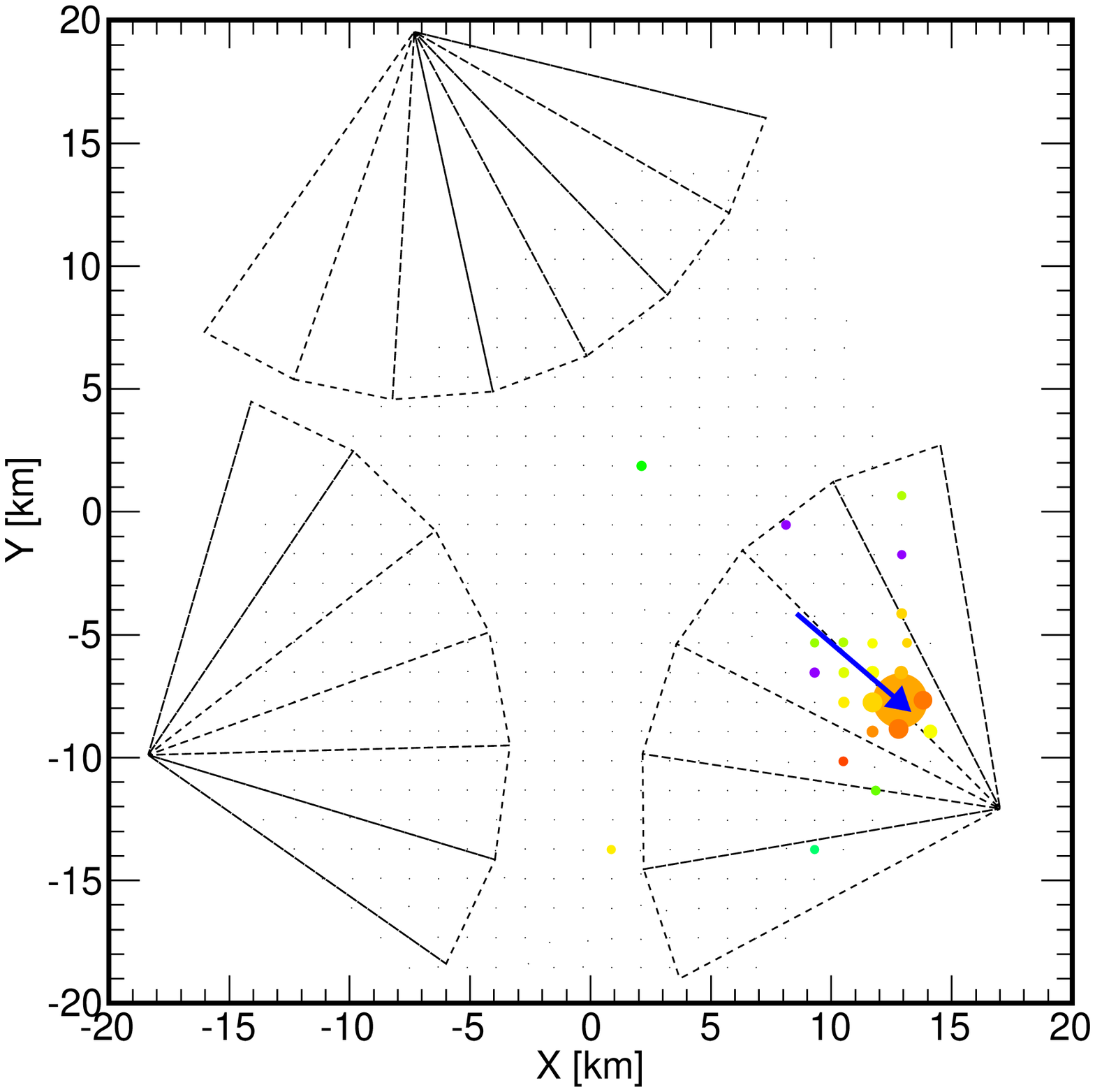}
        \includegraphics[width=0.47\columnwidth]{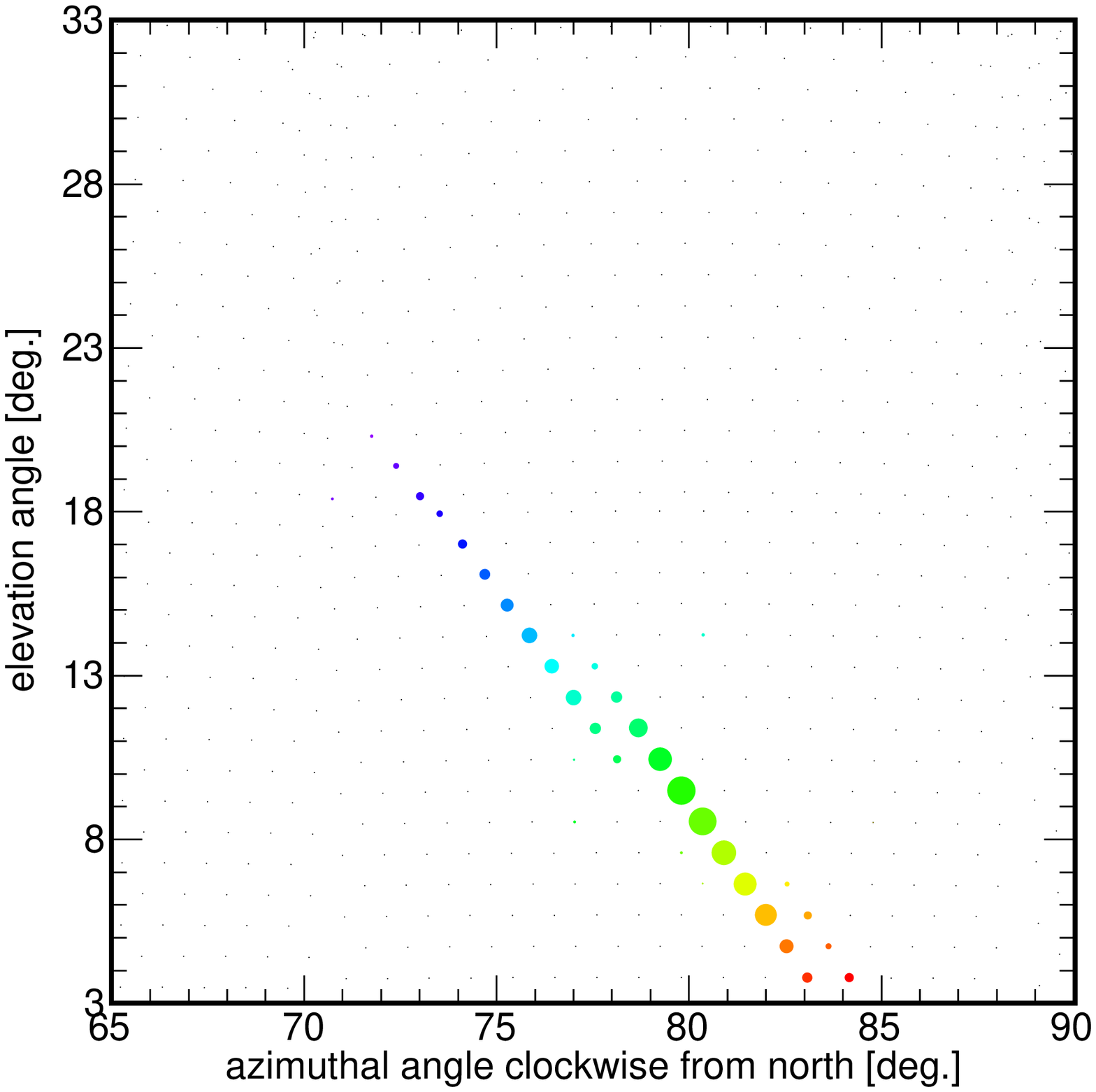}
    \end{center}
    \caption{
    An example event display for a hybrid event. 
    The left figure shows the map of SDs which were hit by the shower. 
    The colors of the filled circles reflect the shower arrival time and the size of the circle is proportional to the number of photo-electrons deposited in the scintillator. 
    The black dotted lines indicate the field of view for the telescopes at each FD.
    The horizontal and vertical axes indicate the locations of the TA detectors, which are the same as in Fig\ref{fig:TA}.
    The blue arrow is the reconstructed shower axis.
    The right figure shows the signals in the LR telescopes.
    The horizontal and vertical axes represent the pointing direction of each PMT.
    The filled circles are the selected PMTs.
    The color indicates timing and the size of the circle indicates the number of detected photo-electrons.}
    \label{fig:event}
\end{figure}

The key idea of the hybrid reconstruction is the use of timing information from one or more SDs in addition to the FD tube timings. The SD timing at which the shower plane crosses the ground gives an ``anchor'' in the conventional FD timing fit and significantly improves the accuracy in shower geometry determination compared to that of the FD monocular mode.
The energy of the UHECR is measured via the calorimetric technique of the FD.
An example of the observed hybrid data is shown in Figure~\ref{fig:event}.

\subsection{PMT Selection}
\label{sec:selection}

The shower analysis procedure begins with selection of PMTs used in the geometry reconstruction among the $256 \times 12$ PMTs in an FD station.
The PMTs to be used are chosen from the ``triggered camera'', in which a shower track is found, and its neighbouring cameras.
First the PMTs with signals greater than $3 \sigma$ above the background fluctuation are selected.
Second the shower track is identified from the PMT hit pattern in the camera(s), and PMTs that are spatially and temporally isolated from the track are rejected.
The bundle of the pointing direction vectors of the PMTs selected at this stage defines the Shower Detector Plane (SDP).
Further selection is made by discarding off-SDP PMTs.
These procedures are iterated until no more PMTs are rejected or reintroduced. 

\subsection{Shower Geometry Reconstruction}
\label{sec:geo-recon}

\begin{figure}[!htbp]
    \begin{center}
        \includegraphics[width=\columnwidth]{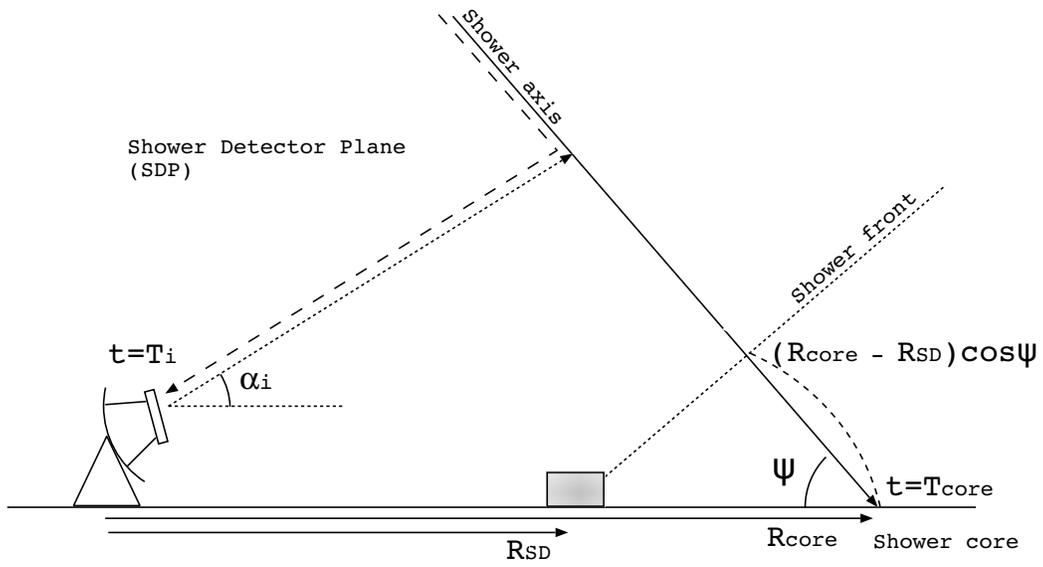}
    \end{center}
    \caption{Diagram indicating the Shower Detector Plane (SDP) used in the time fit.}
    \label{fig:SDP}
\end{figure}

The geometry of the event is determined from the pointing directions and timings of the PMTs of the FD camera:

\begin{equation}
    T_{{\rm exp},i}=T_{\rm core}+\frac{\sin\psi - \sin\alpha_{i}}{c\sin(\psi +\alpha_i)}R_{\rm core},
    \label{eq:FDmonoeq}
\end{equation}
where $T_{{\rm exp},i}$ and $\alpha_i$ are the expected timing and elevation angle in the SDP for the \textit{i}-th PMT, respectively, 
$T_{\rm core}$ is the time when the air shower reached the ground, $R_{\rm core}$ is the distance from the FD station to the core, 
and $\psi$ is the elevation angle of the air shower in the SDP (Figure~\ref{fig:SDP}).

\begin{figure}[!htbp]
    \begin{center}
        \includegraphics[width=\columnwidth]{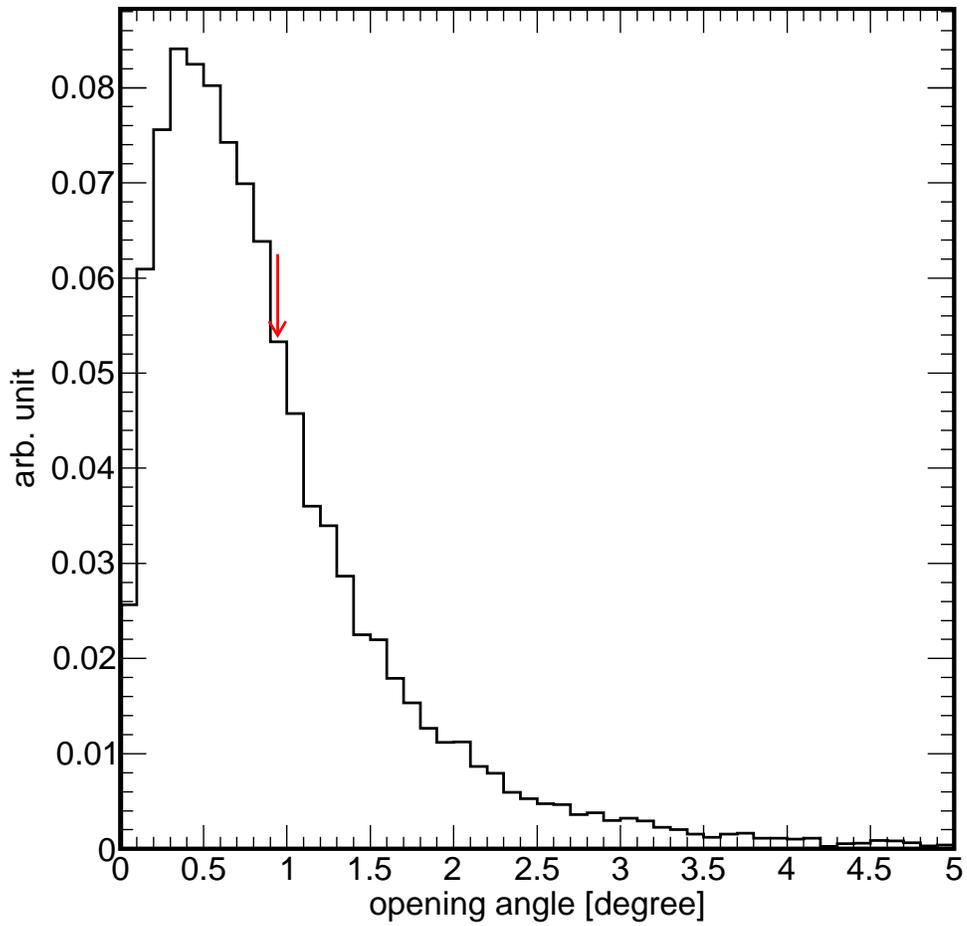}
    \end{center}
    \caption{
    Opening angle between reconstructed and thrown Monte Carlo events.
    Below 0.9 degrees (red arrow), 68.3\% of the reconstructed showers are contained.
    }
    \label{fig:ArrivalDirectionResol}
\end{figure}

For an event that has timing information of  one SD near the core, $T_{\rm core}$ is expressed  by:

\begin{eqnarray}
    T_{\rm core} &=& T^{\prime}_{\rm SD} +  \frac{1}{c}(R_{\rm core}-R_{SD})\cos\psi,\\
    T^{\prime}_{\rm SD} &=& T_{\rm SD} - \frac{1}{c}((\bvec{P^{\prime}}_{\rm SD}-\bvec{P}_{\rm SD})\cdot{\bvec{P}}),
    \label{eq:georeconsd}
\end{eqnarray}
where $\bvec{P}_{\rm SD}$ is the position of the SD, $\bvec{P^\prime}_{\rm SD}$ is the projection of $\bvec{P}_{\rm SD}$ onto the SDP, \bvec{P} is the direction of the shower axis, $T_{\rm SD}$ is the timing of the leading edge of the SD signal.
The quantity to be minimized in the fitting  is written as

\begin{equation}
    \chi^2 = \sum_i\frac{(T_{{\rm exp},i}-T_i)^2}{\sigma^2_{T,i}},
\end{equation}
where $\sigma_{\rm T}$ is the fluctuation of the signal timing.
SDs with distances greater than 1.2km from the line of intersection of the SDP and the ground are rejected, and those farther than 1.5 km from the shower core are also rejected. These procedures are repeated and only one SD that gives the best $\chi^2$ is chosen.
The resolution of the arrival direction is about 0.9~degrees (see Figure~\ref{fig:ArrivalDirectionResol}) which is a significant improvement compared to that in FD monocular mode ($\sim$ 5 degrees).

\subsection{Reconstruction of Longitudinal Shower Profile}
\label{sec:long-recon}

Once the shower geometry is determined, the longitudinal profile of the shower development can be reconstructed from the FD data (the amount of fluorescence photons emitted at various points along the ``known'' shower axis). However there are other components which contribute to the detected signals: $\Cheren$ light beamed near the direction of an air shower, and scattered by atmospheric molecules and aerosols.

In reconstruction of the longitudinal profile, all the detector characteristics including the shadowing effect by the telescope structure, gaps between the mirror segments, the mirror reflectivities, non-uniformities of the PMT cathode sensitivities etc. must be taken into account. This is straightforward in detector simulation using ray-tracing, but not in data reconstruction (for example, it is not possible to know the position at which a photon hit the photo-cathode of a PMT). Therefore we employ an ``inverse MC method'' in shower reconstruction to find an MC shower which best reproduces the data considering all the photon components (fluorescence and $\Cheren$ photons) and detector response.

We assume that the profile of the shower development is represented by the Gaisser-Hillas function~\cite{bib:GH},

\begin{equation}
    N(X; X_{max}, X_0, \Lambda) = N_{\rm max} \left( \frac{X-X_0}{X_{\rm max}-X_0} \right)^{(X_{\rm max}-X_0)/\Lambda}e^{(X_{\rm max}-X)/\Lambda},
    \label{eq:G-H}
\end{equation}
where $X$ is the atmospheric depth, $X_{\rm max}$ is the depth at the shower maximum, $\Lambda$ is the interaction length of the shower particles, and $X_0$ is the offset of $X$. Since $\Lambda$ and $X_0$ do not affect the bulk of the profile, we fix those as 70 g/cm$^{2}$ and 0 g/cm$^{2}$, respectively, and only consider the one parameter $X_{\rm max}$ i.e. $N(X; X_{\rm max})$.

For each air shower event,
the expected number of photo-electrons in the output of the $i$-th PMT in the case of a given $X_{\rm max}$ is obtained by

\begin{eqnarray}
    n_{\rm exp}^i(X_{\rm max}) &=& \sum_k \int_X {\cal N}_k(X; X_{\rm max}) \Phi_k(X) \frac{A(X)}{4 \pi r(X)^2} \epsilon_k(X) \, {\rm d}X \label{eq:nexp},\\
    \epsilon_k(X) &=& S(X) \int_{\lambda} \phi_k(\lambda) T(X, \lambda) R(\lambda) \, {\rm d}\lambda, \label{eq:nexp-epsilon}
\end{eqnarray}
where $k$ is the type of photon production 
(fluorescence light, direct $\Cheren$ light, $\Cheren$ from Rayleigh scattering, and $\Cheren$ from aerosol scattering), 
$X$ is the slant depth along the shower axis, 
${\cal N}_k(X; X_{\rm max})$ is the total number of photons originated at the depth $X$,
$\Phi_k(X)$ is the angular distribution of photons of type $k$ emitted at $X$,
$A(X)$ is the effective mirror area,
and $r$ is the distance between the emission point $X$ to the FD station.
$S(X)$ is the detection sensitivity which includes structure of our telescope and the non-uniformity of photo-cathode surface,
$\lambda$ is the wavelength,
$\phi_k(\lambda)$ is the wavelength spectrum of the process $k$,
$T(X,\lambda)$ is atmospheric transparency,
and $R({\lambda})$ is the detector efficiency. 
Here, atmospheric transparency and detector efficiency are given by

\begin{eqnarray}
    T(X,{\lambda}) &=& T_{\rm Rayleigh}(X,{\lambda})T_{\rm aerosol}(X,{\lambda}),\\
    R({\lambda}) &=& R_{\rm mirror}({\lambda}){\tau}_{\rm filters}({\lambda})P({\lambda}),
\end{eqnarray}
where $T_{\rm Rayleigh}(X,\lambda)$ and $T_{\rm aerosol}$ are the transmittance of the molecular and aerosol atmosphere,
$R_{\rm mirror}(\lambda)$ is the mirror reflectance, $\tau_{\rm f ilters}$ is the transmittance of the ``BG3'' UV-filter and
camera window, and $P({\lambda})$ includes the efficiency of the PMT (quantum efficiency, correction efficiency and gain).

$X_{\rm max}$ is obtained by maximizing the likelihood \textit{L}:

\begin{equation}
    L = \sum_in_{\rm obs}^i \log \left(\frac{n_{\rm exp}^i(X_{\rm max})}{\sum_in_{\rm exp}^i(X_{\rm max})}\right),
\end{equation}
where $n_{\rm obs}^i$ is the sum of the photo-electrons at each PMT, $n_{\rm exp}^i(X_{\rm max})$
is the total number of photo-electrons in the FD station as described in Equation \ref{eq:nexp}.

After fitting for $X_{\rm max}$, $N_{\rm max}$ is obtained by scaling as follows,

\begin{equation}
    N_{\rm max} = \frac{\sum_in_{\rm obs}^i}{\sum_in_{\rm exp}^i(X_{\rm max})}.
\end{equation}

The primary energy is obtained by integration of the Gaisser-Hillas function (Equation \ref{eq:G-H}) with
a correction for the missing energy carried away by neutral particles.

\subsection{Quality Cuts}
To ensure reconstruction quality, we only accept events that satisfy the following criteria:
\begin{itemize}
    \item The number of PMTs used in the reconstruction is greater than 20.
    \item The zenith angle of the reconstructed shower axis is less than 55 degrees.
    \item The shower core is inside the edges of the SD array.  
    \item The angle between the reconstructed shower axis and the telescope is greater than 20 degrees.
    \item $X_{\rm max}$ has to be observed.
\end{itemize}

If events pass the cuts for both the BR and LR stations, we adopt the reconstruction result of the station in which the larger number of PMTs are involved.

\begin{figure}[!htbp]
    \begin{center}
        \includegraphics[width=\columnwidth]{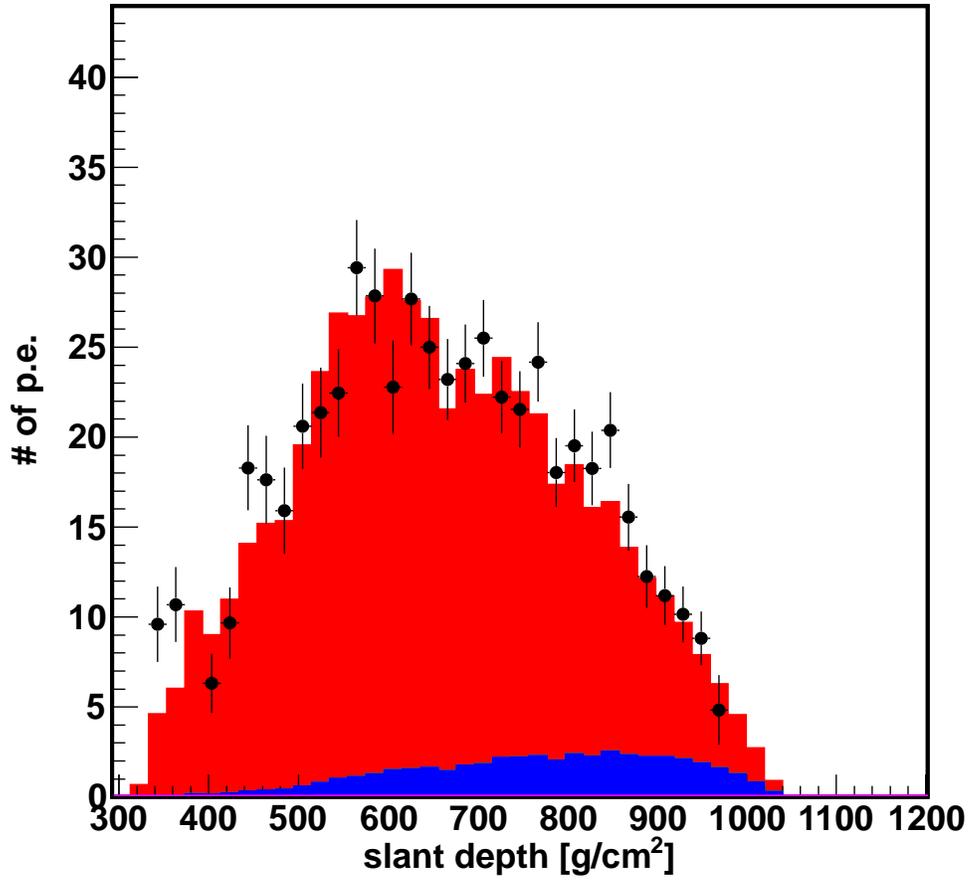}
    \end{center}
    \caption{An example of a reconstruction of the shower profile. The horizontal axis indicates slant depth and vertical axis shows the number of photo-electrons (p.e.) observed by the FD. The black points show the observed data. The filled area represents the fit from the MC event and colors represent the light contribution, red for fluorescence photons, and blue for scattered Cherenkov photons. }
    \label{fig:event-Lng}
\end{figure}

\begin{figure}[!htbp]
    \begin{center}
        \includegraphics[width=\columnwidth]{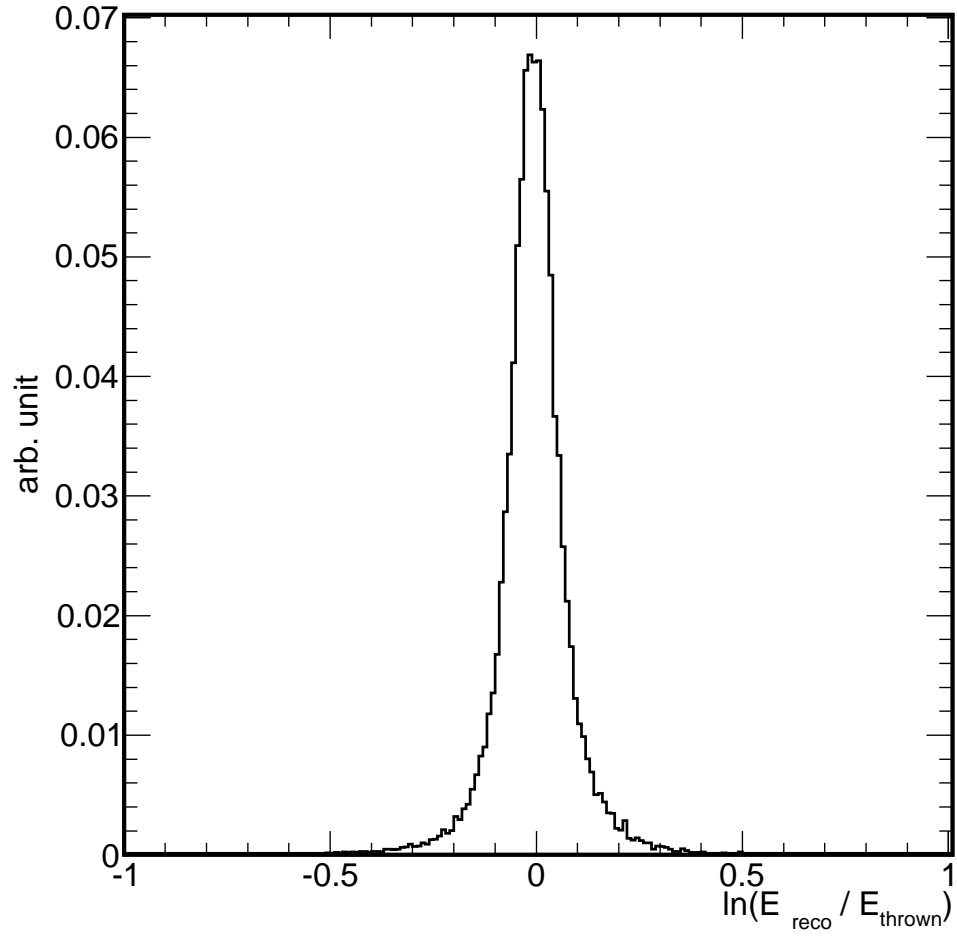}
    \end{center}
    \caption{
    Natural logarithm of the ratio of reconstructed and thrown energies of Monte Carlo simulation events.
    The mean value is 0.0, and 68.3\% of the reconstructed showers are contained within $\pm$0.07.
    }
    \label{fig:EnergyResol}
\end{figure}

An example of the reconstructed shower profile is shown in Figure~\ref{fig:event-Lng}.
For all energy ranges, the energy resolution is on the order of 7\% (see Figure~\ref{fig:EnergyResol}).

\section{Monte-Carlo Simulation of Air Showers and Detectors}
\label{sec:MC}

The performance of our detectors, the reconstruction programs, and the aperture are evaluated using our Monte-Carlo (MC) program. The TA MC package consists of two parts: the air shower generation part and the detector simulation part. In order to reproduce the real observation conditions in the MC, we use environmental data and calibration data that we actually measured at the site assigning a date and time for each MC event. The output of the MC simulation is written out in the same format of the real shower data, so both the MC events and the real shower data can be analysed with the same reconstruction program.

\subsection{Monte-Carlo Simulation of Air Showers}

We generate cosmic-ray showers using the CORSIKA~\cite{bib:CORSIKA} based MC simulation code developed for TA.
The air showers are generated with 10$^{-6}$ thinning to keep fluctuations and event generation times reasonable, and ``dethinned''
to restore the information of individual particles at the ground~\cite{bib:BenMC}.
We use QGSJET-$\rm I\!I$-03~\cite{bib:QGSJET-II} for high energy hadronic interactions and FLUKA-2008.3c~\cite{bib:FLUKA1,bib:FLUKA2} for low energies. Electromagnetic interactions are modeled by EGS4 \cite{bib:EGS4}. We use proton primary particles for the calculation of the aperture. We also use iron to estimate the systematic uncertainty of the aperture.

We generated about 20-million EAS MC simulation events with primary energies ranging from 10$^{17.5}$~eV to 10$^{20.5}$~eV and from 0$^\circ$ to 60$^\circ$ in zenith angle.
For data and MC comparison, the MC events are sampled with the energy spectrum measured by the HiRes experiment~\cite{bib:HiRes-E,bib:HiResx1}, excluding the GZK suppression effect~\cite{bib:GZKg,bib:GZKzk}. A spectral index of 3.25 was used below 10$^{18.65}$ eV and 2.81 above 10$^{18.65}$ eV. 
The positions of the shower cores on the ground were generated within 25~km of the center of the site. The arrival directions are distributed isotropically in the local sky.

\subsection{Monte-Carlo Simulation of Detectors}

The CORSIKA particle outputs (position and momentum of particles at the ground) are used to calculate the energy deposit in each SD with GEANT4~\cite{bib:GEANT4}. The response of the SD electronics is taken into account~\cite{bib:TASD}.
The trigger scheme of the SD array, a three-fold coincidence of adjacent SDs with signals greater than three particle-equivalent, is implemented in the MC. 

The FD simulation includes fluorescence and $\Cheren$ photon generations, telescope optics~\cite{bib:TAFD}, detector calibration~\cite{bib:Tokuno-cal}, and the response of the electronics~\cite{bib:SDF,bib:TFCTD}. 
The CORSIKA output of the longitudinal profile of energy deposit by the charged particles in the atmosphere is used to calculate the number of fluorescence photons emitted at each $1 \, {\rm g/cm^2}$ step. For the fluorescence yield, (the number of photons per energy deposit), we use the value reported by Kakimoto \etal~\cite{bib:KAKIMOTO}. The temperature and pressure dependence of the fluorescence yield is also taken into account by using the radiosonde data~\cite{bib:atm-atten}. The distribution of wavelengths of the fluorescence photons are chosen using the spectrum measured by the FLASH experiment~\cite{bib:FLASH}.

For simulation of $\Cheren$ light emission, we use the energy spectrum of charged particles and angular distribution of produced photons based on CORSIKA~\cite{bib:Nerling}.  We consider $\Cheren$ photons directly detected by the FDs and also scattered photons by molecules and aerosols. A date and time is assigned for each MC event by sampling from the real observation period. The radiosonde data of pressure and temperature as a function of elevation is used to model the molecular atmosphere, and the LIDAR data is used to describe the distribution of aerosols. The measured Vertical Aerosol Optical Depth (VAOD) is 0.035~\cite{bib:atm-atten}.

The telescope simulator includes the segmented mirrors, optical filters, and all obstructions such as camera frames, camera boxes, and shutter frames.
The nightsky background and its fluctuation is taken into account in the simulation by using the mean and variance of the baseline of the PMT outputs recorded in the real data at the assigned time of each MC event.

\subsection{SD Energy Scaling}
From our preparatory study using real shower events detected with both FD and SD, we found that the FD and SD measure the energies of air showers differently. The average of the ratios of the energies independently determined by SD and FD is $\left< E_{\rm SD}/E_{\rm FD} \right> = 1.27$ ~\cite{bib:SDSpec}. Here the energy determination in SD from the particle information at the ground is fully dependent on air shower MC which is based upon hadronic interaction models derived from accelerator experiments in lower energy regions, while the energy can be determined calorimetrically in FD. Therefore we find that an SD reconstruction program tuned by a shower MC like CORSIKA gives $\sim 27\%$ higher energy than the ``true'' energy measured by FD because of the limitations of our present knowledge of air shower phenomena. This difference in the energy scales of FD and SD must be taken into account in the detector simulation and evaluation of the aperture as a function of energy.  

We use a CORSIKA event of energy $E^C$ for detector simulation and aperture evaluation at energy $E = E^C/1.27$, by scaling the longitudinal energy deposit profile of the charged particles in the atmosphere to be measured by FD, and keeping the particle information at the ground and energy deposit in SDs unchanged.  This is simpler than increasing the energy in the SD part, i.e.  the number of particles at the ground and/or the energy deposit in the SDs. We use an elongation rate $d X_{\rm max}/d \log E$ to shift $X_{\rm max}$ in accordance with the $27\%$ energy scaling in the shower profile, but this gives a negligible effect in the energy measurement.

\subsection{Hybrid Aperture and Exposure}
\label{sec:exposure}

The aperture for hybrid events grows with energy, and includes more SDs. However, in the energy region above 10$^{19}$~eV, the aperture for hybrid events saturates since the array edges limit the growth. Thus, the uncertainty of the SD + FD aperture estimation is smaller than that of FD monocular analysis where the aperture continues to grow.  The lower energy bound is given by the efficiency of the SD trigger, a three-fold coincidence of adjacent SDs with signals greater than three particle-equivalent, which falls significantly below $10^{18}$ eV. The typical reconstruction efficiency after all quality cuts is about 70\%. The efficiency is reduced for events with higher energies.  This is caused by the requirement that X$_{\rm max}$ has to be observed within the field of view and the fact that the shower maximum of the events with higher energies sometimes occurs under the ground.

To measure the spectrum with reliable reconstruction, we use data collected on clear and moonless nights with minimal cloud cover in the view of the detector. Weather conditions are recorded for each observation night based on human FD operator's logs. In this analysis, we use 70\% of the total observation time based on the condition that cloud coverage is less than half the sky. The total observation time after subtracting the dead time of the detector is 1480 hours for BR and LR, which consists of 990 hours for stereo observation, 330 hours for BR only and 160 hours for LR only.  

\begin{figure}[!htbp]
    \begin{center}
        \includegraphics[width=\columnwidth]{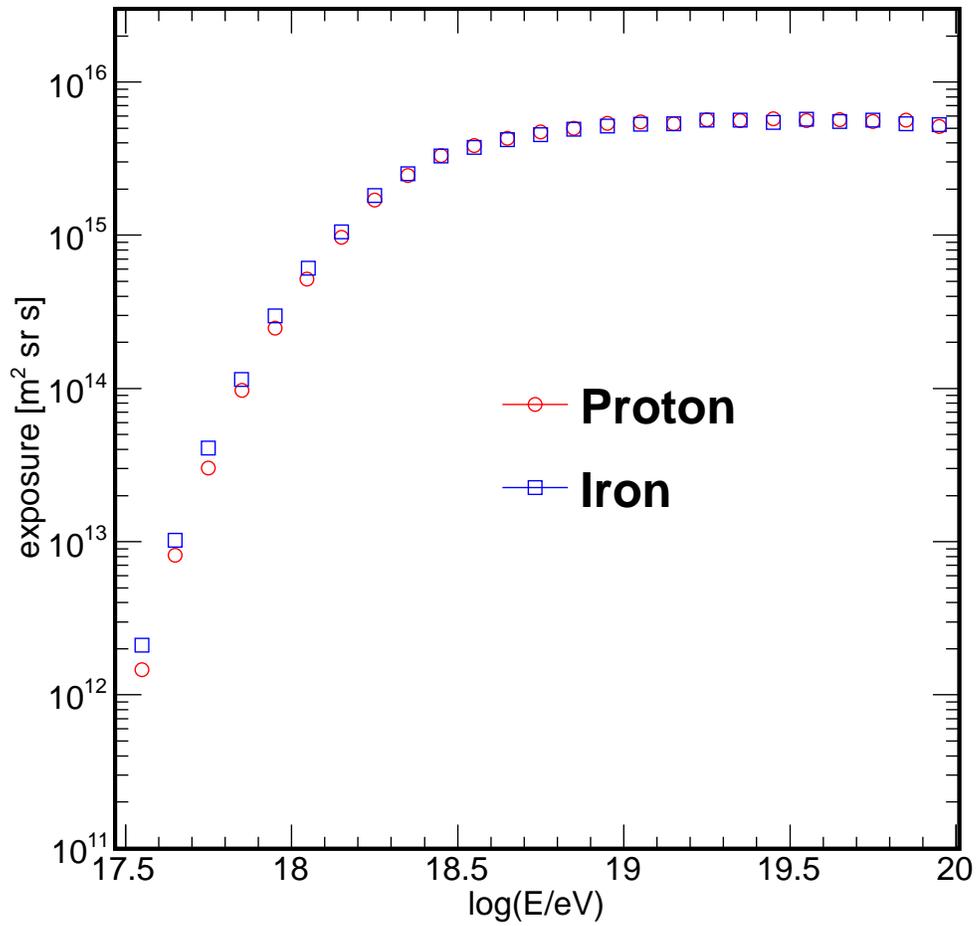}
    \end{center}
    \caption{The calculated hybrid exposure as a function of the energy of the cosmic ray primary. The red circles are proton primaries and the blue squares are iron primaries.}
    \label{fig:HybExposure}
\end{figure}

The aperture of hybrid events with E $>$ 10$^{19}$ eV is 1.2$\times$10$^{9}$m$^2$ sr, which is similar to the SD aperture.
Multiplication by on-time and aperture gives the hybrid exposure for BR and LR. This is calculated to be 6$\times$10$^{15}$ m$^2$~sr~s (Figure~\ref{fig:HybExposure}).

\subsection{Comparison of Data and MC}
\label{sec:comparison}
The quality of the generated MC events is examined by comparing to real data to validate the aperture calculation.
Here we use MC proton and iron showers.

\begin{figure}[!htbp]
    \begin{center}
        \includegraphics[width=\columnwidth]{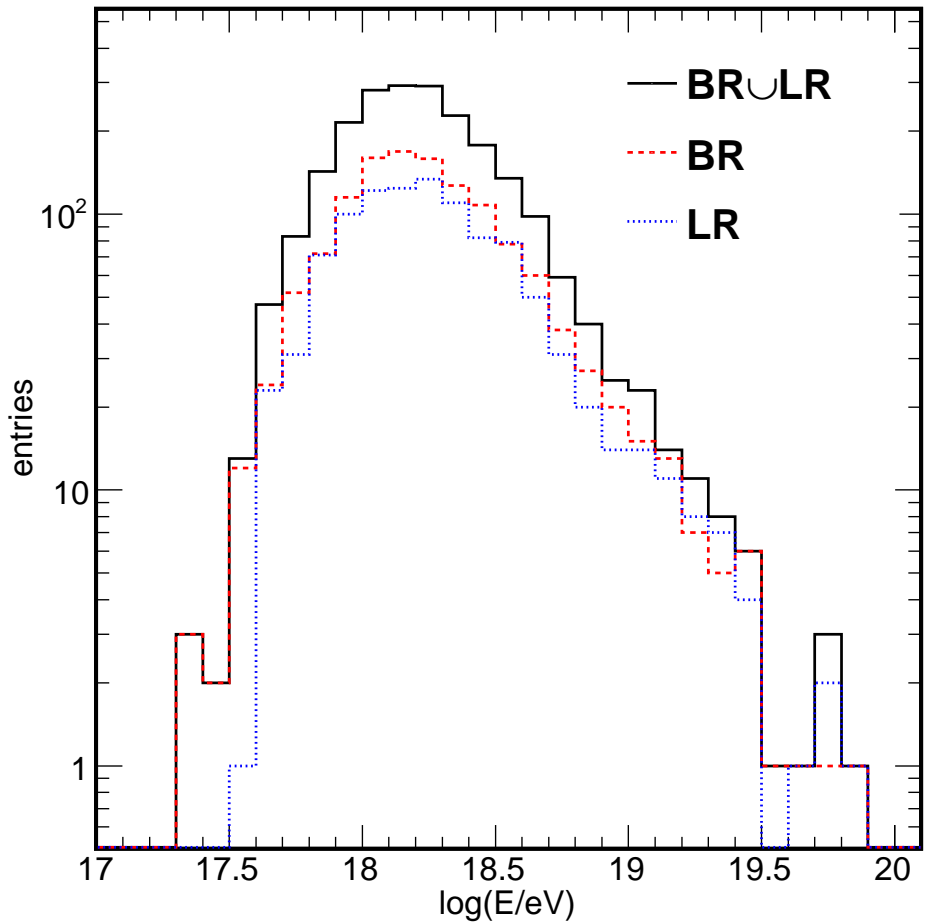}
    \end{center}
    \caption{Raw energy distribution over the first 2.3 years of collection of events observed by the BR and LR fluorescence detectors which coincide with at least one surface detector. The events are shown distributed in tenth-decade energy bins between 10$^{17}$ eV and 10$^{20.1}$ eV.}
    \label{fig:HybEnergy}
\end{figure}

We use shower events detected with the SDs and FDs at the BR and LR sites collected from May 2008 to September 2010.
A total of 3405 events were recorded in the period, and 2203 events remain after hybrid reconstruction and quality cuts (see Section \ref{sec:analysis}). Among the 2203 events, 1276 are from BR and 1040 are from LR, and we find 113 ``stereo'' events that are detected at both BR and LR. The difference in the number of events from the two sites is consistent with the difference in the telescope on-time and the slightly different aperture due to the elevations of the sites and the distance to the closest SDs.
The energy distribution of the observed hybrid events is shown in Figure~\ref{fig:HybEnergy}.

\begin{figure}[!htbp]
    \begin{center}
        \includegraphics[width=\columnwidth]{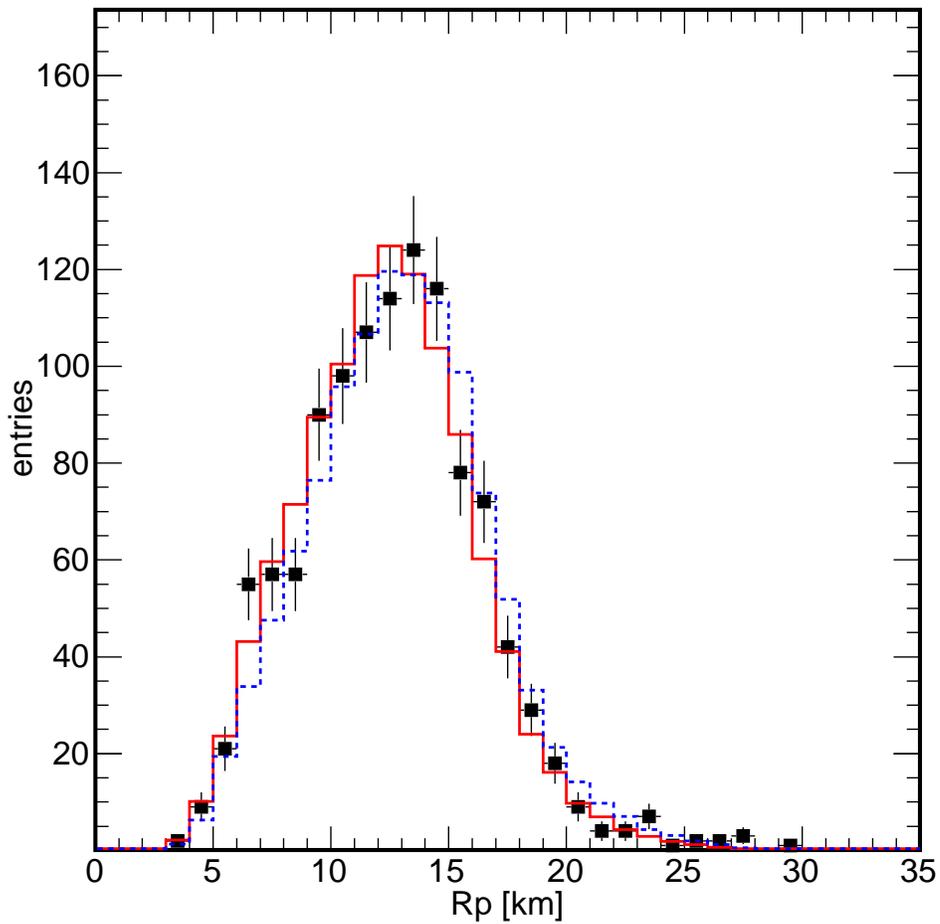}
    \end{center}
    \caption{Comparison of the data and Monte Carlo distribution of the impact parameter, $R_P$. 
    The data is shown by squares with error bars and the Monte Carlo simulation is shown by the histogram.
    The Monte Carlo histogram is normalized to the numbers of data events. The red solid line represents protons and the blue dotted line is iron.}
    \label{fig:RP}
\end{figure}

\begin{figure}[!htbp]
    \begin{center}
        \includegraphics[width=\columnwidth]{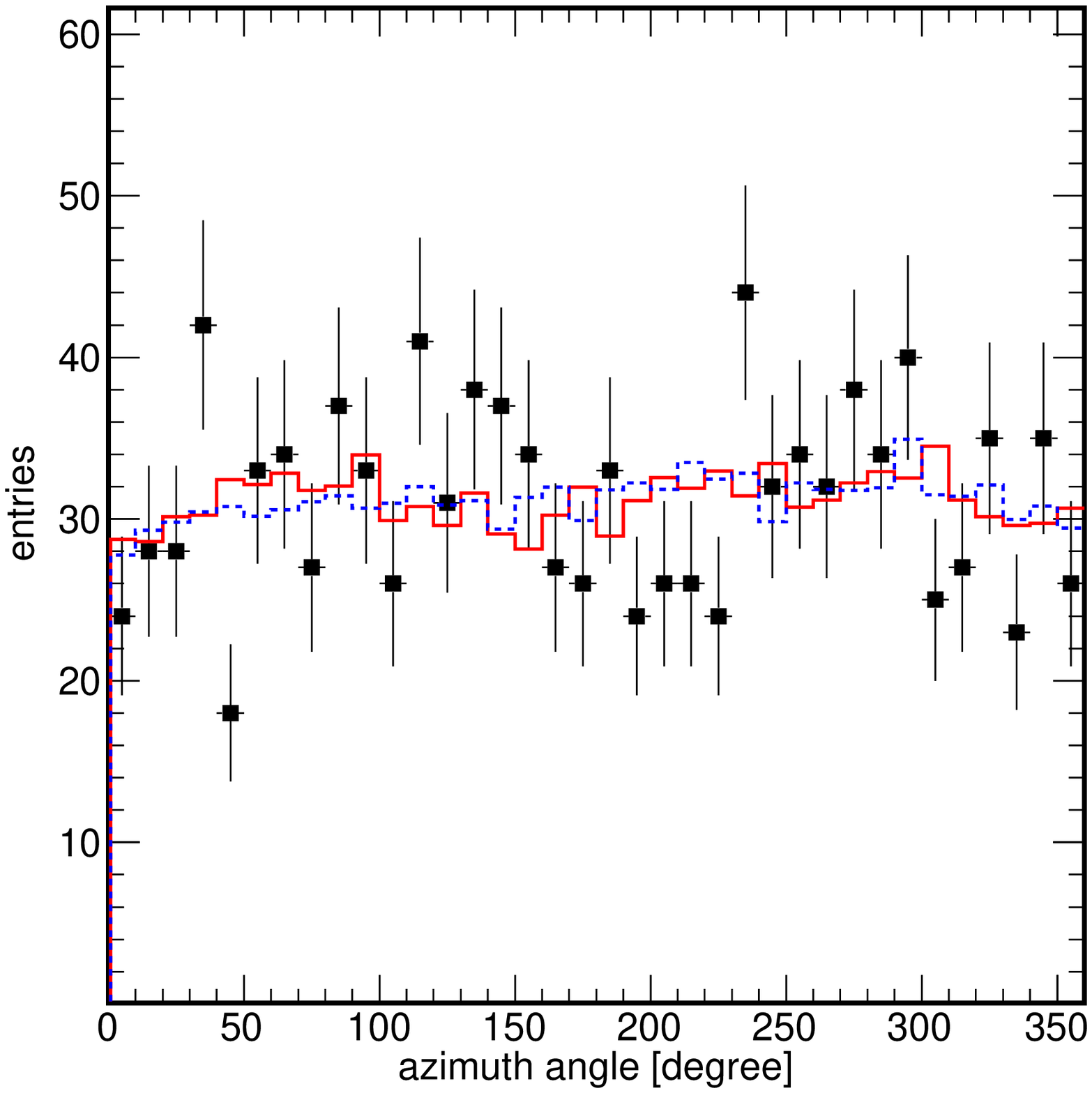}
    \end{center}
    \caption{Comparison of the data and Monte Carlo distributions of the azimuthal angle, $\phi$.
    The data is shown by squares with error bars and the Monte Carlo simulation is shown by the histogram.
    The Monte Carlo histogram is normalized to the numbers of events in the data. The red solid line is proton and the blue dotted line is iron. }
    \label{fig:azimuth}
\end{figure}

Here we show the comparison of MC and the real data in terms of several quantities that are sensitive to the aperture, the shower impact parameter $R_P$, and the shower arrival direction angles $\theta, \phi$ (Figure \ref{fig:RP} $\sim$ \ref{fig:azimuth}). 
For all the parameters, the data and MC events are in excellent agreement.

\section{Result and Discussion}
\label{sec:result}

\begin{figure}[!htbp]
    \begin{center}
        \includegraphics[width=\columnwidth]{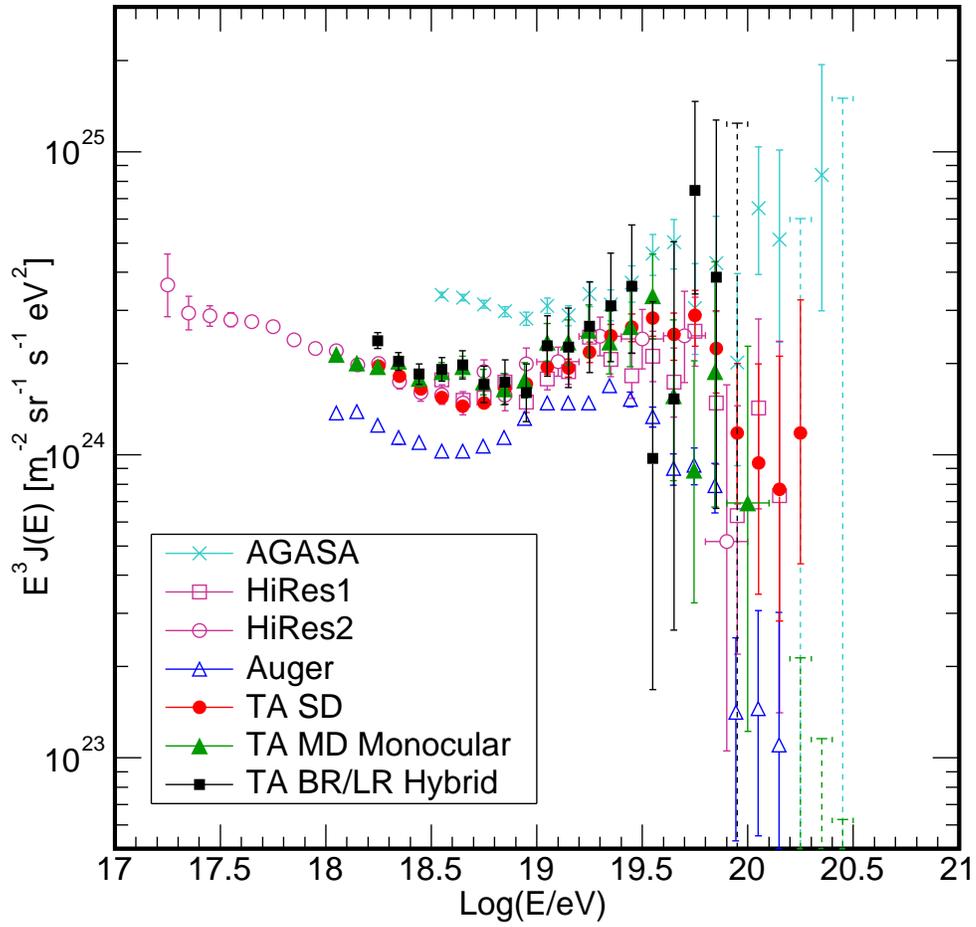}
    \end{center}
    \caption{The energy spectra multiplied by $E^3$. The spectrum determined from the hybrid data is shown by the black boxes.
    The spectra of AGASA~\cite{bib:AGASA-E}, HiRes-1/HiRes-2~\cite{bib:HiRes-E}, Auger~\cite{bib:PAO-E2},TA SD~\cite{bib:SDSpec} and TA MD~\cite{bib:TAMDSpec} are also shown for comparison. }
    \label{fig:spectrum}
\end{figure}

The energy spectrum of cosmic rays, $dI/dE(E)$, is calculated from the number of events in an energy bin and the exposure,

\begin{equation}
    dI/dE(E) = \frac{n(E)}{{\cal E}(E)}
\end{equation}
where $n(E)$ is the number of events in a given energy bin, ${\cal E}(E)$ is the energy-dependent exposure obtained from MC.
Figure~\ref{fig:spectrum} shows the energy spectrum above 10$^{18.2}$~eV.
For comparison, the spectra of AGASA~\cite{bib:AGASA-E}, HiRes~\cite{bib:HiRes-E}, Auger~\cite{bib:PAO-E2}, TA MD~\cite{bib:TAMDSpec} and TA SD~\cite{bib:SDSpec} are also plotted in the same figure.
The TA hybrid spectrum and our previously published spectra are in agreement with HiRes results.

The systematic uncertainties in energy determination are summarized in Table~\ref{tbl:sysE}. 
Systematic uncertainties includes uncertainties in the fluorescence yield (11\%),
atmospheric attenuation (11\%)~\cite{bib:atm-atten},
the absolute detector calibration (10\%)~\cite{bib:Tokuno-cal,bib:calib2,bib:calib3} and
reconstruction (10\%).
The total systematic uncertainty in energy determination is 21\% adding all the uncertainties in quadrature.
This translates to a systematic uncertainty in the flux, $dI/dE$, of 41\% assuming a spectral index of -2.8~\cite{bib:SDSpec}.

A systematic uncertainty in the energy spectrum also comes from the difference in the aperture of the detector to primary cosmic rays of different nuclear types, as shown in Figure \ref{fig:HybExposure}. The difference in the aperture for proton and iron showers increases at lower energies, and amounts to $\sim 10\%$ at $E = 10^{18.2}$ eV, which decreases $dI/dE(E)$ by at most $10\%$ if there are heavier components. 



\section{Summary}
\label{sec:summary}

The Telescope Array including the fluorescence telescopes and the surface detector array has been fully operational since May 2008 . We have developed a hybrid reconstruction technique for air showers using the longitudinal shower profile from FD and the particle arrival timing at the SD. The arrival direction and energy of an air shower can be determined with accuracies of $0.9^{\circ}$ and $7\%$. These are significantly improved compared to FD monocular mode. The systematic uncertainty in determination of energies is evaluated as $21\%$.  

We determine the energy spectrum of cosmic rays with energies above $10^{18.2}$ using the hybrid reconstruction technique using both FD and SD data. The aperture of the detectors is evaluated by taking into account the details of detector performance and atmospheric conditions at the site. The result in this work is in agreement with our previously published spectra obtained from the SD and FD monocular analyses.

\section*{Acknowledgments}
The Telescope Array experiment is supported 
by the Japan Society for the Promotion of Science through
Grants-in-Aid for Scientific Research on Specially Promoted Research (21000002) 
``Extreme Phenomena in the Universe Explored by Highest Energy Cosmic Rays'', 
and the Inter-University Research Program of the Institute for Cosmic Ray 
Research;
by the U.S. National Science Foundation awards PHY-0307098, 
PHY-0601915, PHY-0703893, PHY-0758342, PHY-0848320, PHY-1069280, 
and PHY-1069286 (Utah) and 
PHY-0649681 (Rutgers); 
by the National Research Foundation of Korea 
(2006-0050031, 2007-0056005, 2007-0093860, 2010-0011378, 2010-0028071, R32-10130);
by the Russian Academy of Sciences, RFBR
grants 10-02-01406a and 11-02-01528a (INR),
IISN project No. 4.4509.10 and 
Belgian Science Policy under IUAP VI/11 (ULB).
The foundations of Dr. Ezekiel R. and Edna Wattis Dumke,
Willard L. Eccles and the George S. and Dolores Dore Eccles
all helped with generous donations. 
The State of Utah supported the project through its Economic Development
Board, and the University of Utah through the 
Office of the Vice President for Research. 
The experimental site became available through the cooperation of the 
Utah School and Institutional Trust Lands Administration (SITLA), 
U.S.~Bureau of Land Management and the U.S.~Air Force. 
We also wish to thank the people and the officials of Millard County,
Utah, for their steadfast and warm support. 
We gratefully acknowledge the contributions from the 
technical staffs of our home institutions as well as 
the University of Utah Center for High Performance Computing (CHPC). 




\begin{thebibliography}{00}

\bibitem{bib:TA1} H.~Kawai \etal, J.~Phys.~Soc.~Jpn.~Suppl. A 78 (2009) 108-113.
\bibitem{bib:TA2} H.~Sagawa, Proceedings of the 31st \ICRC, Lodz, Poland (2009). 
\bibitem{bib:TAFD} H.~Tokuno \etal, \NIMPA 676 (2012) 54-65.
\bibitem{bib:SDSpec} T.~Abu-Zayyad \etal, ApJ 768 L1 (2013).
\bibitem{bib:HiResMIA} T.~Abu-Zayyad \etal, \PRL 84 (2000) 4276-4279.
\bibitem{bib:PAO-E} J.~Abraham \etal, \PRL 101 (2008) 061101.
\bibitem{bib:TASD} T.~Abu-Zayyad \etal, \NIMPA 689 (2012) 87-97.
\bibitem{bib:TAMDSpec} T.~Abu-Zayyad \etal, \APP 109 (2012) 39-40.
\bibitem{bib:SDF} A.~Taketa \etal, Proceedings of the 29th \ICRC, Pune, India (2005).
\bibitem{bib:TFCTD} Y.~Tameda \etal, \NIMPA 609 (2009) 227-234.
\bibitem{bib:atm-atten} T.~Tomida \etal, \NIMPA 654 (2011) 653-660.
\bibitem{bib:GH} T.K.~Gaisser and A.M.~Hillas, Proceedings of 15th \ICRC, Plovdiv, Bulgaria (1977).
\bibitem{bib:CORSIKA} D.~Heck, G.~Schatz, T.~Thouw, J.~Knapp, and J.N.~Capdevielle, Tech. Rep. 6019, FZKA (1998).
\bibitem{bib:BenMC} B.T.~Stokes \etal, \APP 35 (2012) 759-766.
\bibitem{bib:QGSJET-II} S.~Ostapchenko, Nucl. Phys. Proc. Suppl. 151 (2006) 143-146. 
\bibitem{bib:FLUKA1} A.~Ferrari, P.R.~Sala, A.~Fasso, and J.~Ranft, Tech. Rep. 2005-010, CERN, (2005).
\bibitem{bib:FLUKA2} G.~ Battistoni \etal, AIP Conf. Proc. 896 (2007) 31-49.
\bibitem{bib:EGS4} W.R. Nelson, H. Hirayama, D.W.O. Rogers, Tech. Rep. 0265, SLAC (1985).
\bibitem{bib:HiRes-E} R.U.~Abbasi \etal, \PRL 100 (2008) 101101.
\bibitem{bib:HiResx1} R.U.~Abbasi \etal, \PRL 104 (2010) 161101.
\bibitem{bib:GZKg} K.~Greisen, \PRL 16 (1966) 748-750.
\bibitem{bib:GZKzk} G.T.~Zatsepin and V.A.~Kuz'min, JETP Lett. 4 (1966) 78-80.
\bibitem{bib:GEANT4} S.Agostinelli \etal, \NIMPA 506 (2003) 250-303.
\bibitem{bib:Tokuno-cal} H.~Tokuno \etal, \NIMPA 601 (2009) 364-371.
\bibitem{bib:KAKIMOTO} F.~Kakimoto \etal, \NIMPA 372 (1996) 527-533.
\bibitem{bib:FLASH} R.U.~Abassi \etal, \APP 29 (2008) 77-86.
\bibitem{bib:Nerling} F.~Nerling \etal, \APP 24 (2006) 421-437.
\bibitem{bib:AGASA-E} M.~Takeda \etal, \APP 19 (2003) 447-462.
\bibitem{bib:PAO-E2} P.~Abreu \etal, arXiv:1107.4809.
\bibitem{bib:calib2} D.~Ikeda \etal, Proceedings of the 31st \ICRC, Lodz, Poland (2009).
\bibitem{bib:calib3} S.~Kawana \etal, \NIMPA 681 (2012) 68-77.





\end{thebibliography}



\newpage

\newpage

\newpage

\newpage

\newpage

\newpage

\newpage

\newpage

\newpage

\newpage
\begin{figure}[!htbp]
    \begin{center}
        \includegraphics[width=\columnwidth]{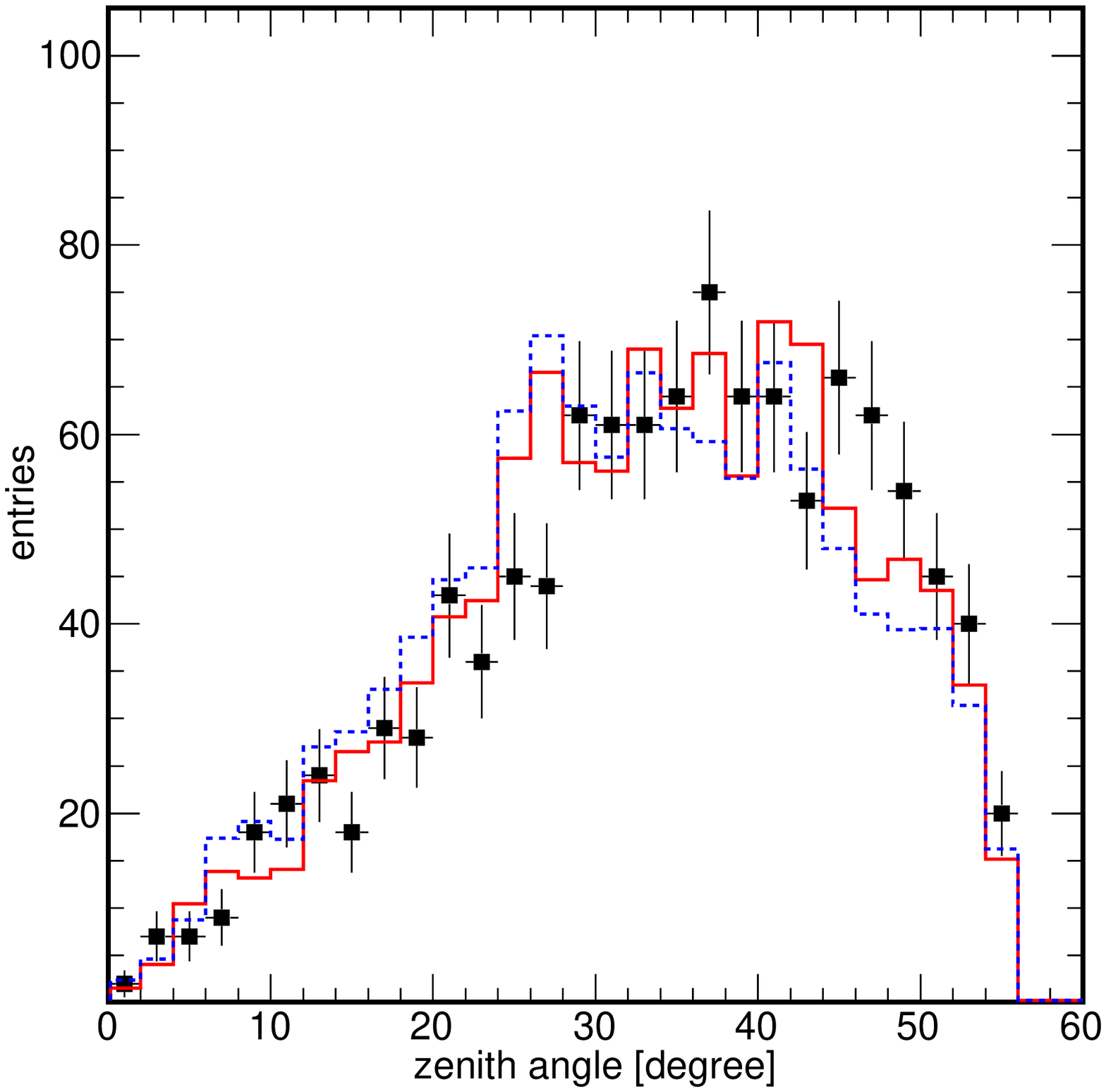}
    \end{center}
    \caption{Comparison of data and Monte Carlo distributions for the zenith angle, $\theta$.
    The data is shown by squares with error bars and the Monte Carlo simulation is shown by the histogram.
    The Monte Carlo histogram is normalized to the numbers of events in the data.  The red solid line is proton and the blue dotted line is iron.}
    \label{fig:zenith}
\end{figure}

\newpage

\newpage


\newpage
\begin{table}[!htbp]
    \begin{center}
        \begin{tabular}{c|c|c} \hline
            Item & Error & Contributions \\ \hline
            Detector sensitivity & 10\% & PMT (8\%), mirror (4\%), \\
            &           & aging (3\%), filter (1\%)\\
            Atmospheric collection & 11\% & aerosol (10\%), \\
            &          & Rayleigh (5\%) \\
            Fluorescence yield        & 11\% & model (10\%), \\
            & & humidity (4\%), \\
            & & atmosphere (3\%) \\
            reconstruction               & 10\% & model ( 9\%) \\ 
            &           & missing energy (5\%) \\ \hline
            Sum in quadrature            & 21\% & \\ \hline
        \end{tabular}
    \end{center}
    \caption{Systematic uncertainties of energy measurement.}
    \label{tbl:sysE}
\end{table}

\end{document}